\documentclass{revtex4}

\usepackage{amsmath}
\usepackage{amsfonts}
\usepackage{amssymb}
\usepackage{bbold}

\newcommand{\bra}[1]{\left\langle #1 \right|}
\newcommand{\ket}[1]{\left| #1 \right\rangle}
\newcommand{\scalar}[2]{\left\langle #1 | #2 \right\rangle}
\newcommand{\vect}[1]{\mathbf{#1}}
\newcommand{\average}[1]{\left\langle #1 \right\rangle}

\newcommand{\boldpi}{\boldsymbol{\pi}}
\newcommand{\boldepsilon}{\boldsymbol{\epsilon}}
\newcommand{\R}{\mathbb{R}}
\newcommand{\C}{\mathbb{C}}

\begin{document}

\title{Multiple classical limits in relativistic and nonrelativistic quantum mechanics}
\author{N.\ Yokomizo}
\author{J.\ C.\ A.\ Barata}
\affiliation{Instituto de F\'{i}sica, 
Depto.\ de F\'{i}sica Matem\'{a}tica, 
Universidade de S\~{a}o Paulo, C.P.\ 66318, 05315-970 S\~{a}o Paulo-SP, Brazil}
\date{\today}

\begin{abstract}

  The existence of a classical limit describing interacting particles
  in a second-quantized theory of identical particles with bosonic
  symmetry is proved. This limit exists in addition to a previously
  established classical limit with a classical field behavior, showing
  that the limit $\hbar \rightarrow 0$ of the theory is not unique. An
  analogous result is valid for a free massive scalar field: two
  distinct classical limits are proved to exist, describing a system
  of particles or a classical field. The introduction of local
  operators in order to represent kinematical properties of interest
  is shown to break the permutation symmetry under some localizability
  conditions, allowing the study of individual particle properties.

\end{abstract}

\maketitle

\section{Introduction}

Quantum Field Theory was born in the attempt to conciliate Quantum
Mechanics and Relativity and in the attempt to deal with the so-called
``particle-wave duality'' underlying quantum phenomena by making the
corpuscular character of matter compatible with the classical notion
of field. Its main purpose is to describe particle physics, but the
way in which its fundamental principles admit or recognize the notion
of particle is rather indirect, if not obscure. A fundamental attempt
to understand when quantum field theories describe particles was the
work of Haag and Swieca \cite{haag-swieca}, further developed by
Buchholz and Wichmann in \cite{buchholz-wichmann} and in subsequent works. In a nutshell, these works point to the fact that
any relativistic quantum field theory describing particles must have
some specific limitations on the number of degrees of freedom at
finite volume and limited energy.

Quantum Field Theory is believed to be, in some sense, a fundamental
theory, but the notions of particles and of fields are derived from
our sensorial experiences in a classical macroscopic world. It is,
therefore, of great importance to have a precise understanding on how
the classical scenario can be reached from that more fundamental
quantum starting point. In particular, one should naturally expect
that the particle-field duality manifests itself in any general
attempt to reach the classical limit of quantum fields. The existence
of these two different limits (particles or fields) lies deeply in the
structure of quantum field theory and its physical interpretation. It
was first remarked in the fundamental work of Hepp \cite{hepp} on the
classical limit of quantum systems and the main purpose of the present
work is to clarify certain aspects of this remark, specially in the
relativistic regime. We believe that the analysis of these multiple
classical limits could have some conceptual importance in the context
of quantum field theories formulated in curved spacetimes, where no
natural concept of particle states is available.

There are several ways to approach the formulation of a classical
limit of Quantum Mechanics. In this work we follow the ideas
introduced by Hepp in \cite{hepp}, which can be applied to a wide
range of systems and can be understood in a simple and precise
way. The central result of his work combines an old observation of
Schr\"odinger \cite{schrodinger}, which led to the discovery of
coherent states \cite{sudarshan,glauber}, with the intuitive explicit
content of Ehrenfest theorem. Schr\"odinger observed that in a
harmonic oscillator a Gaussian wavefunction moves without distortion
along a classical orbit, what led him to try to understand particles
in general as localized wavepackets in motion. But that could not be
accomplished, since wavepackets in general spread, so that particles
cannot remain localized indefinitely. On the other hand, Ehrenfest
theorem describes the time-dependence of quantum averages with
classical-looking equations. The main drawback is that for a function
of the position operator $A(q)$ one has $\average{A(q)} \neq
A(\average{q})$, unless $A$ is linear. In the case where $A$ is an
external force, if one considers $\average{q}$ as representing the
position of the particle, a force $A(\average{q})$ would be expect to
act on it, whereas by Ehrenfest theorem the force is
$\average{A(q)}$. Thus $\average{q}$ deviates from the classical
trajectory.

In order to briefly describe Hepp's analysis, consider a simple
example. The classical nonrelativistic motion of a single particle of
mass $m$ moving in one-dimension under an external potential $V$ is
described in Classical Mechanics by the Hamiltonian $ \mathsf{H} =
\frac{p^2}{2m} + V(x) $, leading to the canonical equations $ \dot{p}
= - \frac{\partial V}{\partial x}$, $\dot{q} = \frac{p}{m} $.  Let
$(\xi(t), \pi(t))$ be the solution for initial conditions $x(0) = \xi,
\, p(0) = \pi$. The dynamics of the corresponding quantum system is
defined by the Hamilton operator
\begin{equation}
H = - \frac{\hbar^2}{2m} \frac{\partial^2}{\partial q^2} + V(q) \, ,
\label{particle-hamiltonian}
\end{equation}
where $q$ is the position operator. Let $U(t)$ be the propagator
associated to this Hamiltonian. The question we face is how to recover
the classical trajectory defined by $\xi(t)$ and $\pi(t)$ from the
quantum Hamiltonian. Following Hepp, there are three steps involved in
the solution of this problem. In the first step we go to the Weyl
representation, replacing position and momentum operators by their
exponentiated versions $\exp(i a q)$, $\exp(i b p)$, $a,\, b \in
\R$, so that only bounded operators are involved. In the
second step we consider averages of general Weyl operators
$\mathcal{W}(a, \, b):=\textrm{e}^{i (aq + bp)} $ on time-evolved
coherent states $\ket{\alpha}$, with $\alpha = (\xi + i \pi)/(\sqrt{2
  \hbar})$.  Finally, the limit $\hbar \rightarrow 0$ is taken. As
established in \cite{hepp}, it turns out that under natural regularity
requirements, and for $t$ restricted to a finite interval $|t| < T$,
one has
\begin{equation}
\lim_{\hbar \rightarrow 0} 
\bra{\alpha} U(t)^* \mathcal{W}(a, \, b) U(t) \ket{\alpha} = 
\textrm{e}^{i [a \xi(t) + b \pi(t)]} 
\, ,
\label{one-particle-limit}
\end{equation}
from which the classical trajectory $(\xi(t),\,\pi(t))$ can be
recovered (for the above mentioned time-interval). The result is valid
for a large class of potentials, as stated in \cite{hepp}.

The steps and results leading to \eqref{one-particle-limit} can be
easily extended in order to indicate how the classical limit of a
quantum nonrelativistic system with finitely many degrees of freedom
describing distinguishable particles has to be performed in order that
this limit describes a classical mechanical system of finitely many
particles (at least for a short time). In this case one can use
coherent states as those of Eq.~\eqref{one-particle-limit} in
connection with individual position and momentum operators $p_i,\,
q_i$ for each particle to find the classical limit of interacting
point particles. The procedure presented above have also been extended
in \cite{hepp} to some quantum systems with infinitely many degrees of
freedom, leading to classical limits describing classical fields.

An important quantum system considered in \cite{hepp} for which the
classical limit is a classical field is a a system of second-quantized
interacting bosonic particles moving in one dimension, described in a
bosonic Fock space $\mathcal{F}_s(\mathcal{H}) = \oplus_{n=0}^\infty
\mathcal{H}_s^{(n)}$, where $\mathcal{H}_s^{(n)}$ is the usual Hilbert
space of symmetric square-integrable functions over $\mathbf{R}^n$,
with the Hamiltonian
\begin{equation}
H 
= 
- \frac{\hbar^2}{2} \int \textrm{d}x \, a^*(x) \nabla^2 a(x) 
+ \frac{1}{2} \int \textrm{d}x \textrm{d}y \, a^*(x) a^*(y) V(x-y) a(x) a(y)
 \, ,
\label{eq:fock-hamiltonian}
\end{equation}
where the creation and annihilation operators $a^*$ and $a$ satisfy
the usual commutation rules,
\[
[a(x),\, a(y)] = [a^*(x),\, a^*(y)] = 0 
\, , \qquad 
[a(x),\, a^*(y)] = \delta(x-y) 
\, ,
\]
and $V(x)=V(-x)=V(x)^*$ is some real Kato potential. The classical
limit of this system was studied in \cite{gross} as a model for
superfluid Helium, and that work was one of the motivations for
Hepp's. In this case the classical field behavior is taken as an
approximate description for the dynamics of the matter density in the
fluid. That limit can be achieved with the same strategy adopted for
the case of the one-particle dynamics of
\eqref{particle-hamiltonian}. Coherent states of bosonic systems are
usually defined in order to make field aspects become more evident, as
can be seen from many examples in Quantum Optics (see e.g.\
\cite{glauber63b} for an introduction). Accordingly, if one considers
the time-evolution of the average of exponentiated smeared local
fields evaluated on an initially coherent state, one gets a classical
field theory in the limit $\hbar \rightarrow 0$. As shown in
\cite{hepp}, the classical field $\alpha$ satisfies the non-linear
partial integro-differential equation
\begin{equation}
\frac{\partial \alpha}{\partial t}(\beta,t,x) 
\;=\; 
\frac{i}{2 \mu} \nabla^2 \alpha(\beta,t,x) 
+ i \int \textrm{d}y \, V(x-y) |\alpha(\beta,t,y)|^2 \alpha(\beta,t,x) 
\, ,
\label{eq:field-limit}
\end{equation}
where $\beta(x)$ is the initial condition (which must be taken in
$D(\nabla^2)$), and the constant $\mu$ is related to the mass $m$ of
the bosonic particles. The particular case when $V(x-y)=g\delta(x-y)$
leads in \eqref{eq:field-limit} to the well-known Gross--Pitaevskii
equation (or non-linear Schr\"{o}dinger equation), widely employed in
the study of Bose-Einstein condensates. The field behavior exposed in
\eqref{eq:field-limit} is not expected to hold along an arbitrary time
interval.

Since \eqref{eq:fock-hamiltonian} is assumed to describe a quantum
system of interacting particles, it is natural to expect that a second
kind of classical limit exists which describes classical particles
instead of classical fields. In fact, Hepp observes that these limits
should depend on the way in which certain physical parameters are
scaled when $\hbar\to 0$. There are, however, some implicit
difficulties in applying Hepp's program to systems of
indistinguishable particles, as in the case of the nonrelativistic
quantum many-body system described in \eqref{eq:fock-hamiltonian} or a
relativistic quantum field model. The main problems are that : (i) the
coherent states are not invariant under permutation symmetry, and (ii)
observables describing individual kinematical properties may not be
defined. The problem (i) is circumvented with the use of symmetrized
products of single-particle coherent states, but the second problem
requires some variation of the general technique. We will show how the
introduction of local operators acting on a class of essentially
localized states defined herein can be used to address this problem. A
classical limit of $N$ interacting particles is found when $N$
apparatuses situated at distinct regions are used to observe an
essentially localized state with $N$ localization centers coinciding
with the apparatuses' positions.

In the relativistic regime, one has to face the additional problem of
defining the notion of quantum particle-like states in a proper way,
since this is required for a particle classical limit. We adopt the
notion of essentially localized states discussed by Haag and Swieca in
\cite{haag-swieca} as a suitable representation of the intuitive idea of a
particle, and adjust it to our purposes. Position averages will be
evaluated with the Newton-Wigner operator \cite{newton-wigner}. An
explicit construction based on the single-particle relativistic
coherent states of \cite{kaiser} will be shown to lead to the desired
particle classical limit of a free massive scalar field. As an
intermediate step, we discuss the classical limit of the
first-quantized Klein-Gordon field and show that the methods of
\cite{hepp} can be adapted to solve this problem.  Since the field
classical limit of this theory is proved in \cite{hepp}, we conclude
again that the classical limit is not unique.

\section{Coherent states and essentially localized states}

\label{sec:coherent}

Coherent states $\ket{\alpha}$ for a one-dimensional quantum
mechanical system are usually defined as eigenstates of the
annihilation operator, $a \ket{\alpha} = \alpha \ket{\alpha}$, with
$\alpha = (\xi + i \pi)/\sqrt{2 \hbar} \in \C$. It follows
from this definition that $\ket{\alpha}$ is a minimum uncertainty
state centered at $x=\xi$ and $p=\pi$ with equal uncertainties $\Delta
p = \Delta q$, and this property is one of the main motivations for
the study of such states, since it allows one to think of them as the
``most classical states'' in some natural sense. An alternative and
equivalent definition is given in terms of the action of a translation
operator $U(\alpha)$ on the harmonic oscillator ground state
$\ket{0}$,
\begin{equation}
\ket{\alpha} := U(\alpha) \ket{0} 
\, , \qquad 
U(\alpha) := \exp(\alpha a^* -\bar{\alpha} a) 
\, , 
\label{eq:coherent-number}
\end{equation}
where one has $U(\alpha) a U(\alpha)^* = a - \alpha$. The scalar
products are given by
\begin{equation}
|\scalar{\alpha}{\beta}|^2 = \exp (-|\alpha - \beta|^2) \, ,
\label{eq:overlap-squared}
\end{equation}
thus coherent states are not orthogonal; nevertheless, the overlap
decreases rapidly with the distance $|\alpha - \beta|$.

Coherent states can be defined in much more general systems (see
\cite{gilmore, perelomov}). The case of $n$-dimensional systems is
straightforward, as well as the case of many-particles systems when no
symmetrization is required. Let $d$ be the number of spatial
dimensions, and $n$ the number of particles. In this case the
coordinate space is $N$-dimensional, with $N=nd$. All that is needed
is to define a series of $N$ annihilation operators $a_r = (q_r + i p_
r)/{\sqrt{2 \hbar}}$, one for each spatial dimension and particle of
the system, and consider the simultaneous eigenstates of all
annihilation operators, $a_r \ket{\alpha_1,\, \dots,\, \alpha_N} =
\alpha_r \ket{\alpha_1,\, \dots,\, \alpha_N}$, $r=1,\, \dots,\,
N$. The eigenstates $\ket{\alpha_1,\, \dots,\, \alpha_N}$ are the
coherent states for this system. Writing the labels $\alpha$ in terms
of real and imaginary parts as $\alpha_r = (\xi_r + i \pi_r) / \sqrt{2
  \hbar}$, one finds that $\ket{\alpha_1,\, \dots,\, \alpha_N}$
describes a Gaussian wavefunction centered at $\vect{x}=(\xi_1,\,
\dots,\, \xi_N)$ in coordinate space and $\vect{p} = (\pi_1,\,
\dots,\, \pi_N)$ in momentum space with minimum uncertainty for each
canonical pair, $\Delta q_r = \Delta p_r = \sqrt{\hbar/2}$. So
everything goes as in the previous case.

The case of many-body systems of identical particles, as the one
described by the Hamiltonian \eqref{eq:fock-hamiltonian} or in Quantum
Field Theory, demands special care. Consider an $N$-particle state of
a many-body system of identical particles in one space dimension and
with bosonic symmetry (to which we restrict ourselves in this
work). Because of the permutation symmetry, one cannot introduce
individual position and momentum operators for each of the particles
as in the case of distinguishable particles and a different procedure
is required. The most natural choice, and the one adopted in this
work, is to consider symmetrized products of single-particle coherent
states $\ket{\alpha_i}$,
\begin{equation}
\ket{\alpha_1,\, \dots,\, \alpha_N}_S 
\, = \,  
\frac{1}{\mathcal{N}} \sum_\pi \ket{\alpha_{\pi(1)},\, \dots,\, \alpha_{\pi(N)}} 
\, = \,  
\frac{1}{\mathcal{N}} \sum_\pi \ket{\alpha_{\pi(1)}}\otimes\cdots\otimes\ket{\alpha_{\pi(N)}} 
\, ,
\label{eq:n-particle-coherent}
\end{equation}
as the analog of $N$-particle coherent states, where the sum is over
all elements of the permutation group of $N$ elements and where the
normalization constant is
\begin{equation}
\mathcal{N} \, =\, 
 N! \sum_\pi \scalar{\alpha_1}{\alpha_{\pi(1)} } \, \cdots \, \scalar{\alpha_N}{\alpha_{\pi(N)}}  
\;.
\end{equation}
These states will be of special interest in what follows, as examples
of what will be called ``essentially localized states''. It will be
argued that such states reproduce within nonrelativistic quantum
mechanics some basic intuitive properties of states so denoted in
Algebraic Quantum Field Theory \cite{haag-swieca}.

In order to discuss localization properties consider the example of
one-dimensional single-particle states first, with the Hilbert space
$\mathcal{H}=L^2(\R, \, dx)$.  Denote by $\langle A
\rangle_\psi := \big\langle \psi |A|\psi\big\rangle$ the expectation
value of an operator $A$ acting on $\mathcal{H}$ in some vector state
for $\psi\equiv\ket{\psi}\in \mathcal{H}$ with $\|\psi\|^2
\equiv\scalar{\psi}{\psi}=1$.

Let $A$ be a bounded operator acting on $\mathcal{H}$ with norm $\| A
\|$. For any finite open region $O\subset \R$, define the
local version of $A$ corresponding to the region $O$ by
\begin{equation}
A_O \; := \; \frac{1}{2} \big( \chi_O A + A \chi_O  \big) \;,
\label{eq:Definicao-de-AO}
\end{equation}
where $ \chi_O $ is the characteristic function of $O$, i.e.,
$\chi_O(x)=1$ if $x\in O$ and $\chi_O(x)=0$ if $x\not\in O$.  The
operator $A_O$ is also a bounded operator acting on $\mathcal{H}$ and
one has $A-A_O=\frac{1}{2}\big[(1-\chi_O) A + A
(1-\chi_O)\big]$. Hence, for expectation values of $ A-A_O$ on some
normalized vector state $\psi$ one has
\[
\langle A \rangle_\psi - \langle A_O \rangle_\psi
\, = \, 
\frac{1}{2}\big\langle (1-\chi_O)\psi\big|  A \big|\psi\big\rangle
+
\frac{1}{2}\big\langle \psi\big|  A \big|(1-\chi_O)\psi\big\rangle
\, ,
\]
from which we get, by the Cauchy-Schwarz inequality,
\[
\big| \langle A \rangle_\psi - \langle A_O \rangle_\psi \big|
\, \leq \, 
\|A\|\, \big\| (1-\chi_O)  \psi\big\|
\; .
\]

Let us consider the concrete case where $O=(\xi-R,\, \xi+R)$, the
interval of radius $R>0$ around the center $\xi\in\R$. We say
that $\psi\in\mathcal{H}$ is essentially localized around $\xi$ if $\|
\psi \|_{O^c} := \|(1-\chi_O) \psi\|$, the fraction of the norm of the
state lying outside $O$, satisfies
\begin{equation}
\| \psi \|_{O^c}
\, := \,
\|(1-\chi_O)  \psi\|
\, \leq \, 
\mathcal{K}\, \textrm{e}^{- R^2 / 2 \hbar}
\, ,
\label{eq:localized-1state-bound}
\end{equation}
for all $R>0$, $\mathcal{K}$ being some positive constant. For vector
states satisfying \eqref{eq:localized-1state-bound}, one has
\begin{equation}
|\langle A \rangle_\psi - \langle A_O \rangle_\psi |
\, \leq \,
\mathcal{K}\,\| A \| \textrm{e}^{- R^2 / 2\hbar} \, . 
\label{local-approx-av}
\end{equation}
Thus the operator $A_O$ can in fact be understood as a local version
of the operator $A$ in the region $O$, since the expectation $\langle
A_O \rangle_\psi$ is a good approximation for $\langle A
\rangle_\psi$, up to an error which decreases rapidly with $R$.

In particular, this approximation works fine for coherent states.  It
was already mentioned that the position wavefunction
$\psi_\alpha(x)=\scalar{x}{\alpha}$, corresponding to a coherent state
$\ket{\alpha}$ is a normalized Gaussian curve with width
$\sqrt{\hbar}$ centered around some $\xi\in\R$. Thus
$\psi_\alpha(x)$ is essentially contained in a radius of a few
$\sqrt{\hbar}$ around $\xi$ and one has
\begin{equation}
\| \psi_\alpha \|_{O^c}^2
\, := \,
\|(1-\chi_O)  \psi_\alpha\|^2 
\, = \, 
 1 - \int_{\xi - R}^{\xi + R} |\psi_\alpha(x)|^2 \textrm{d}x 
\, \leq \, 
\mathcal{K}\textrm{e}^{- R^2 /  \hbar} 
\, ,
\label{eq:localized-1state}
\end{equation}
for some $\mathcal{K}>0$, and decreases faster than exponentially with increasing $R$.

There is another sense in which \eqref{eq:localized-1state-bound}
leads to a notion of essentially localized states. Consider two
distinct states $\ket{\psi}$ and $\ket{\phi}$ which satisfy
\eqref{eq:localized-1state-bound} with respect to disjoint regions $O$
and $O^\prime$, and let the distance between such regions be
$d(O,O^\prime) = d > 0$. Hence, $ |\scalar{\psi}{\phi}| \leq 2
\textrm{e}^{- d^2/8\hbar} $. Therefore, if $\ket{\psi}$ and
$\ket{\phi}$ are states essentially localized in disjoint regions, the
overlap between the corresponding wavefunctions is small for large
distances, and decreases as a Gaussian with the distance between the
localization regions. Accordingly, the inequality
\eqref{eq:localized-1state-bound} is used herefrom as the defining
property of a ``state essentially localized in $O$''.

The concepts of local operator and essentially localized states can be
easily translated to $N$-particle states. The symmetrized product of
single-particle states $\ket{\psi_i}$ essentially localized in
pairwise disjoint regions $O_i$ (with $O_i \cap O_j = \emptyset$ if $i
\neq j$) is the natural extension adopted here. These states will be
denoted ``multiply localized states'' or ``essentially localized
states''. Simple examples are given by the symmetrized product of
coherent states as displayed in Eq.~\eqref{eq:n-particle-coherent}, as
long as the localization centers $\xi_i$ of the coherent states
$\ket{\alpha_i}$ are chosen sufficiently far apart. Given a local
operator $A_O$ acting on the single-particle subspace we define its
$N$-particle counterpart acting on $\mathcal{H}_s^{(N)}$ as
\begin{equation}
A^{(N)}_O = \sum_{i=1}^N \mathbb{1} \otimes \cdots \mathbb{1} \otimes A_O \otimes \mathbb{1} \otimes \cdots \otimes \mathbb{1}
\, ,
\label{eq:nbody-local-op-def}
\end{equation}
where the operator $A_O$ occupies the $i$-th slot of the tensor
product in the $i$-th term of the sum.

The average value of the local operator $A^{(N)}_O$ on a multiply
localized state $\ket{\psi}$ is given by
\begin{equation}
\left\langle A^{(N)}_O \right\rangle_\psi 
= 
\sum_i \sum_\pi \frac{N!}{\mathcal{N}^2}
\scalar{\psi_1}{\psi_{\pi(1)}} \cdots \bra{\psi_i} A_O\ket{\psi_{\pi(i)}} \cdots \scalar{\psi_N}{\psi_{\pi(N)}} 
\, .
\label{eq:nbody-local-operator-psi}
\end{equation}
The factor $N! / \mathcal{N}^2$ goes to $1$ when $\hbar \rightarrow
0$. In the case where $\ket{\psi}$ is a multiply localized state with
localization centers $O_i$, one has
\begin{equation}
\scalar{\psi_i}{\psi_{\pi(i)}} 
\; \leq \; 
2 \textrm{e}^{- d^2/8\hbar} 
\, , \qquad 
\textrm{ if } \, i\neq \pi(i) \, , 
\label{scalar-product-bound}
\end{equation}
where $d$ is the smallest distance between any two $O_i,\;O_j$, $i\neq
j$. Thus, only a few terms in \eqref{eq:nbody-local-operator-psi}
contribute appreciably to the sum, i.e., those for which $i=\pi(i)$,
$\forall i$. Each such terms contains a factor $\bra{\psi_i} A_O
\ket{\psi_i}$. Now, suppose some of the localization centers coincides
with the region where the local operator in question is defined: $O_k
= O$ for some $k$. Then one has the bounds
\begin{eqnarray}
|\langle A \rangle_{\psi_k} - \langle A_O \rangle_{\psi_k} | & \leq &  \| A \| \textrm{e}^{- R^2 / \hbar} \, , 
\\
| \langle A_O \rangle_{\psi_i} | & \leq & \| A \|  \textrm{e}^{- d^2/ \hbar} \, , \qquad \textrm{if } i\neq k \, .
\label{local-op-bounds}
\end{eqnarray}
We assume in addition that
\begin{equation}
| \bra{\psi_i} A_O \ket{\psi_j} | \,  \leq \,  \| A \| \textrm{e}^{-  d^2/2 \hbar} \, , \qquad \textrm{if } i\neq j \, .
\label{local-op-cond}
\end{equation}
This inequality corresponds to the requirement of a local property of
the operators. It is automatically satisfied when $A_O$ is the local
version of any operator $A$ of the form $A= A(q)$. It is also
satisfied when $A= A(q) \exp(i b p / \hbar)$, with $b \leq d$, that
is, when the operator $A$ is a function of the coordinates, up to some
small translation. In this case, we get from
\eqref{eq:nbody-local-operator-psi} the approximation
\begin{equation}
\left\langle A^{(N)}_O \right\rangle _\psi 
\; \approx \; 
\langle A \rangle_{\psi_k} 
\, ,
\label{eq:op-approximation}
\end{equation}
with an error which decreases as $\exp(- d^2/4 \hbar) \exp(- R^2/
\hbar)$ for large arguments $d,\, R$. Therefore, the average value of
a local many-body operator associated with a region $O$, evaluated on
a multiply localized state which has $O$ as one of its localization
centers, reduces to the average value of the corresponding single
particle operator evaluated on a state essentially localized above
$O$.

The intuitive picture underlying the definitions and approximations
given in this section is the following. In a many-body system of
identical particles, one cannot in general disentangle kinematical
properties of individual components, since permutation symmetry mixes
them. But for some special states, an approximation is feasible where
the particles behave as independent subsystems. Such ``essentially
localized states'' look like isolated lumps of matter distributed over
distinct regions of space. If a measuring apparatus works in a region
where no lump of matter is present it will detect nothing. If there is
one lump of matter present, a single particle will be detected, but
the observer will be unaware of the existence of other identical
particles comprising a larger Hilbert space together with the detected
one --- his ``local operators'' break the permutation symmetry since
they are not sensible to particles far apart. If the measuring
apparatus operates in a region where several lumps of matter are
present, the individual properties of the particles become
intertwined. The configurations of interest for the existence of a
classical limit describing particles are those in which $N$ observers
situated at distinct regions can measure properties of one particle
each.

\section{Classical limit of interacting bosonic particles}

\label{sec:nonrelat-particles}

A vector state in $\mathcal{H}^{\otimes N}_s$ can be represented by a
normalized wavefunction $\psi(x_1, \dots, x_N)$ symmetric under the
exchange of any two coordinates $x_i,\, x_j$. Let $\ket{\alpha_i}$ be
single-particle coherent states and $\psi_i(x)$ be the corresponding
wavefunctions. The symmetrized product of such states is represented
by
\begin{equation}
\psi(x_1,\dots, x_N) = \frac{1}{\mathcal{N}} \sum_{\pi} \psi_{\pi(1)}(x_1) \dots \psi_{\pi(N)}(x_N) \, ,
\label{eq:initial-state}
\end{equation}
where $\mathcal{N}$ is a normalization constant. The action of the
Hamiltonian \eqref{eq:fock-hamiltonian} on such states is given by
\begin{multline}
H \psi(x_1,\dots, x_N) 
= 
- \frac{\hbar^2}{2} \frac{1}{N!} \sum_j \sum_{\pi} \nabla^2_j \psi_{\pi(1)}(x_1) \dots \psi_{\pi(N)}(x_N) 
\\
+ 
\frac{1}{2} \frac{1}{N!} 
\sum_{j \neq k} \sum_{\pi} V(x_j - x_k) \psi_{\pi(1)}(x_1) \dots \psi_{\pi(N)}(x_N) 
\, .
\label{wavefunction-hamiltonian}
\end{multline}
Since the Hamiltonian is time-independent, the propagator is simply
$U(t) = \exp (i H t/\hbar)$. The potential $V(x)$ is required to
satisfy the regularity condition $\int \textrm{d}x \, |V(x)|^2 \exp(-
\rho x^2) < \infty$ for some $\rho < \infty$ \cite{hepp}. This quantum
dynamics is to be compared with the classical Hamiltonian equations
\begin{gather}
\mathsf{H}_c = \sum_j \frac{\pi_j^2}{2} + \frac{1}{2} \sum_{j \neq k} V(\xi_j - \xi_k) \, , 
\\
\dot{\xi_j} =  \pi_j \, , \qquad  \dot{\pi_j} = - \sum_{k \neq j} V^\prime(\xi_j - \xi_k) \, .
\label{classical-equations}
\end{gather}
Following \cite{hepp}, we assumed that $\nabla^2 V$ is Lipschitz, so
that solutions of the canonical equations exist and are unique in a
finite time-interval $|t| \leq T$. Denote by $\xi(\alpha,t), \,
\pi(\alpha,t)$ the solution for initial conditions $\alpha_j = (\xi_j
+ i \pi_j)/\sqrt{2 \hbar}$, $j=1, \, \ldots , \, N$, and let $V$ be
H\"older continuous $C^{2+\epsilon}$ in an open neighborhood of
$\xi(\alpha,t)$, for all $|t| \leq T$.

In order to compare the classical and quantum equations, consider for
each $j$ the localized operators $\mathcal{W}(a, \, b)_{O_j(t)}$
associated to the Weyl operators $\mathcal{W}(a, \, b):=\exp i (a q +
b p) $, acting on the single particle space, given as in
\eqref{eq:Definicao-de-AO} by \[ \mathcal{W}(a, \, b)_{O_j(t)} \; :=
\; \frac{1}{2}\Big(\chi_{O_j(t)} \exp i (aq + b p) + \exp i (a q + b
p) \chi_{O_j(t)}\Big) \; ,
\]
where $O_j(t) := (\xi_j(\alpha,t) - R, \; \xi_j(\alpha,t) + R)$, where
$\chi_{O_j(t)}$ is the characteristic function of the set $O_j(t)$,
and $q, \, p$ are the usual position and momentum operators.  For each
$j$, let $\mathcal{W}(a, \, b)_{O_j(t)}^{(N)}$ be the corresponding
$N$-body operator defined as in \eqref{eq:nbody-local-op-def}. The
desired classical limit is encoded in the expression
\begin{equation}
\lim_{\hbar \rightarrow 0} 
\left[
\bra{\psi} U(t)^* \, \mathcal{W}(a, \, b)_{O_j(t)}^{(N)}\, U(t) \ket{\psi}
-
\textrm{e}^{i [a \xi_j(t) +  b\pi_j(t) ]} 
\right]
\; = \; 0
\; ,
\label{eq:classical-limit}
\end{equation}
valid for each $j=1, \, \ldots , \, N$, where $\ket{\psi}$ is the
state defined in \eqref{eq:initial-state}.  This limit is proved as
follows.

Consider the left side of the equation \eqref{eq:classical-limit}. The
state $\ket{\psi}$ is a superposition of $N!$ unsymmetrized
coherent-states in $\mathcal{H}^{\otimes N}$. Let $\ket{\phi}$ be one
of these states. The Hamiltonian \eqref{wavefunction-hamiltonian} can
be understood as acting on $\mathcal{H}^{\otimes N}$, and in this case
it is known from Hepp's work \cite{hepp} that, under the assumed
hypotheses,
\[
\lim_{\hbar \rightarrow 0} \Big\| U(t)\ket{\phi} - U(\alpha(t)) W(t)\ket{0} \Big\| 
= 
0 \, ,
\]
where $W(t)= \text{T} \exp \left[ - i / \hbar \int_0^t
  \text{d}t^\prime H^\prime(t^\prime) \right]$ (here, $\text{T}$
denotes the usual time-ordering prescription) is the propagator
associated with the second-order Hamiltonian
\[
H^\prime 
\; = \; 
- \frac{\hbar^2}{2} \sum_j \nabla^2_j 
+ \frac{1}{2} \sum_{j \neq k} V^{\prime \prime} (\xi_j(t) - \xi_k(t)) (x_j - x_k)^2
\;,
\]
obtained from the linearization of the operator in
\eqref{wavefunction-hamiltonian} around the classical orbit
$\xi(\alpha,\,t), \, \pi(\alpha,\, t)$. The state $\ket{0}$ is the
coherent state centered at zero position and momentum. To the state
$W(t)\ket{0}$ there corresponds a wavefunction which is Gaussian and
centered at zero in each coordinate $x_i$ (an explicit expression is
derived in Appendix \ref{time-evolution}). Thus one can write
\begin{gather}
\phi^\prime(x,\, t) = \bra{x} U(\alpha(t)) W(t)\ket{0} = \prod_j
\phi_j^\prime(x_j,\, t) \, 
\\ \qquad 
\phi_j^\prime(x_j,\, t) 
= 
\left( \frac{\omega_j}{\pi \hbar} \right)^{1/4} 
\exp \left[ 
   - \frac{1}{2 \hbar} \omega_j(t) (x_j - \xi_j(\alpha,\, t))^2  
   + i \frac{\pi_j(\alpha,\, t)}{\hbar} x_j 
\right] 
\, .
\label{phi-prime}
\end{gather}
The positive coefficients $\omega_i(t)$ are continuous functions on
$[0,\, T]$ determined by the Hamiltonian $H^\prime$. The state $U(t)
\ket{\psi}$ is the symmetric part of $U(t) \ket{\phi}$, which for
small $\hbar$ is well approximated by
\begin{equation}
\psi(x,\, t) 
\, \approx \, 
\frac{1}{N!} 
\sum_\pi \phi^\prime_{\pi(1)}(x_1,\,t) \dots \phi^\prime_{\pi(N)}(x_N,\, t) 
\, ,
\label{eq:psi-t}
\end{equation}
with an error of order $O(\hbar^{\delta / 2})$. Define $\Omega := \min
\big\{ \omega_j(t)| \,j=1,\, \dots,\, N \text{ and } t \in[0,\,
T]\big\}$. Then \eqref{phi-prime} ensures that the states
\eqref{eq:psi-t} are essentially localized around the classical orbit
$\xi(\alpha,t)$, i.e., $ \| \phi_j^\prime(x,\, t) \|_{O_j^c(t)} \leq
\textrm{e}^{- R^2 / 2 \hbar^\prime} $, with $\hbar^\prime = \hbar /
\Omega$.

Now observe that the Weyl operators can be rewritten as
$\mathcal{W}(a, \, b):=\exp(i ab \hbar/2 )\exp(i a q) \exp(i (b \hbar)
p/ \hbar)$, so there is some $\hbar$ such that $b \hbar$ is less than
the smallest distance between the particles. Thus the inequality
\eqref{local-op-cond} is valid, and then the approximation
\eqref{eq:op-approximation} yields, for each $k=1, \, \ldots , \, N$,
\[
\bra{\psi} U(t)^* \; \mathcal{W}(a, \, b)_{O_k(t)}^{(N)} \; U(t) \ket{\psi} 
\; \approx \;
 \bra{\phi^\prime_k(t)} \,\mathcal{W}(a, \, b)\, \ket{\phi^\prime_k(t)} 
\, .
\]
This approximation is valid as long as the classical orbits
$\xi_j(\alpha,\, t)$, $j=1, \, \ldots , \, N$, do not cross, i.e., $|
\xi_j(\alpha,\, t) - \xi_l(\alpha,\, t)| > 2R + \kappa$, for all $j
\neq l$ and $t \in [0,\, T]$, $\kappa > 0$, in order that a finite
minimal distance $\kappa$ exists between the localization regions
centers $)_j(\alpha,\, t)$. The error involved in the approximation
vanishes for $\hbar \rightarrow 0$, and the average values on the
right side can be easily evaluated, leading to the limit
\eqref{eq:classical-limit}, and completing the proof.

Some comments are in order. First, it is clear that there is a lot of
freedom in the choice of the local operators, since there are many
ways to define localization regions $O_j(t)$. In the proof displayed
above, these regions follow the classical trajectory with some fixed
radius $R$. This is not necessary --- it is sufficient that the
classical trajectory of the $j$-th particle is in the interior of
$O_j(t)$. If there are disjoint regions $O_j$ such that $\xi(t) \in
O_j$ for all $t$, then the time-dependence can be removed. In this
case one would have a physical situation where the measuring
apparatuses are placed at fixed regions where the particles are
confined, a situation which is likely to happen for small
time-intervals. Another remark concerns the possibility of collisions
between particles. In this case the classical limit does not exist in
the sense of the proved theorem, so the classical limit of scattering
processes are not considered here.

Regarding the existence of two distinct classical limits in the same
quantum system, it is seen that the existence of two kinds of coherent
states for bosonic systems is responsible for the fact. The usual
field coherent states lead to a classical field equation obeying
\eqref{eq:field-limit} for a time-interval of order $t \sim \hbar^2$,
while the essentially localized states defined herein lead to a
particle dynamics for a time-interval of order $\hbar^{\delta/2}$. The
introduction of local operators is necessary in the case of the
particle limit in order to break the permutation symmetry and allow
for the study of individual trajectories.

A natural question connected with Hepp's analysis is to what extent it
helps understanding the existence of a `classical world' as a
consequence of more fundamental quantum laws. In short, what it states
is that under a certain condition -- the existence of a coherent state
-- and for a small time-interval, operator averages obey the expected
classical laws of motion. Now, this result raises two natural
questions. The first is to understand why a classical behavior is
usually met with in macroscopic scales, that is, why states other than
coherent states usually do not show up in macroscopic scales. Another
question concerns the possibility of removing the restriction to small
time-intervals, in order that this classical limit exists in
time-scales compatible with all classical dynamics.

A possible improvement of the theorem in order to address these
problems could be the inclusion of an external system interacting with
the system of interest. That would bring some contact with the widely
studied decoherence approach to the emergence of classical behavior in
quantum systems \cite{zurek}. It is known that several new effects can
show up in this case, such as Zeno effect \cite{itano} and non-unitary
time-evolution \cite{lindblad}, for instance, and it is not
unreasonable that they may play an important role in the existence of
a classical limit for large time-scales. Some experimental evidence
points in this direction. A clean example is found in a series of
papers on interference of matter waves (fullerenes) in a Talbot-Lau
interferometer \cite{zeilinger1,zeilinger2}. It was observed that both
interaction with gas particles as well as emission of radiation, i.e.,
interaction with a quantized electromagnetic field, helps preserving
classical behavior. In both cases a process of localization of the
particles is supposed to happen with a certain frequency in
consequence of the interaction with the external system, thus
naturally enforcing the regular occurrence of the conditions required
for the validity of Hepp's analysis.

\section{Classical particles and the Klein-Gordon field}

\label{sec:relat-particles}

In this section we extend our results to the relativistic regime. We
will discuss the existence of two distinct kinds of classical limits
in a system of bosonic relativistic particles, one of them describing
a classical field, the other one describing classical systems with
finitely many particles. For simplicity, we consider a scalar field
theory with mass $m>0$ in $1+1$-dimensional Minkowski spacetime. The
existence of a classical field limit for this system in the presence
of a polynomial interaction was proved in \cite{hepp}, roughly in the
same way as for nonrelativistic mechanics: one studies the
time-evolution of average values evaluated in an initially coherent
state, and verifies that these averages obey the expected classical
equations of motion. Here we show that a classical particle limit can
also be reached, using methods analog to those applied in the
nonrelativistic case, i.e., by studying the time-evolution of the
average of local operators evaluated on essentially localized
states. This will be done explicitly for the case of the relativistic
position operator, in order to show how classical trajectories arise
from the quantum dynamics.

It is well-known that the problem of localizability of relativistic
particles is rather more intricate than in the nonrelativistic
case. The basic difference is that here the construction of a
wavepacket cannot involve arbitrary superpositions of states, being
restricted to the space of positive energy solutions. It turns out
that a strictly localized state cannot be constructed, in contrast
with the nonrelativistic case where arbitrarily localized states can
be easily written down. A relativistic particle is at best ``nearly''
localized, i.e., concentrated in a small region of space. The
questions of how well localized the particle can be, and how to
characterize these localized states, were dealt with in the classical
work of Newton and Wigner \cite{newton-wigner}, where a set of natural
conditions were stated which any localized state should satisfy. The
solutions for these conditions are the so-called Newton-Wigner
states. These states can be characterized as infinite norm eigenstates
of a relativistic version of the position operator. In the following,
we adopt this notion of localizability and use the Newton-Wigner
operator to evaluate position averages where necessary.

The existence of a particular classical limit relies on the existence
of an adequate kind of coherent state. So, in order to formulate a
classical particle limit, one must first look for the analogous of
particle-like coherent states in relativistic quantum theory. This
problem was addressed in several previous works and we refrain from
giving a complete list of references here. The states introduced by
G.\ Kaiser in \cite{kaiser} (which are also particular cases of the
formalism developed by S.\ Ali \textit{et alii} in \cite{ali}) proved
to be a good starting point for our purposes. In order to simplify the
calculations, we actually work with a simple modification of these
states. Before we introduce the coherent states we will deal with,
let us recall some aspects of the theory of Newton and Wigner.

Let $\varphi(x,\, t)$ be a classical scalar field satisfying the
Klein-Gordon equation with mass $m$ in $1+1$-dimensional Minkowski
spacetime, and $\phi(p,\, \omega)$ its momentum space representation,
and assume that $\phi(p,\, \omega) \equiv \phi(p)$ is restricted to
the positive mass shell, with $\omega = \sqrt{p^2 + m^2}$ (we adopt
$c=1$ throughout the paper). One can view $\varphi(x,\, t)$ as a first
quantized particle wavefunction or as a state in the one-particle
sector of the Fock space of the corresponding quantum field theory.

In this context, there are two relevant Hilbert spaces to be
considered: $\mathcal{H}^1=L^2(\R, dp/\omega)$, with the
relativistically invariant scalar product
$\scalar{\phi}{\psi}\equiv\scalar{\phi}{\psi}_{\mathcal{H}^1}
:=\int_{\R}\frac{dp}{\omega}\,\overline{\phi(p)}\psi(p)$,
and $\mathcal{H}^2=L^2(\R, \;dp)$, with the usual scalar
product $\scalar{\phi}{\psi}_{\mathcal{H}^2}:=\int_{\R}
dp\,\overline{\phi(p)}\psi(p)$ (both scalar products
interpreted in momentum space representation). The map
$M_{\sqrt{\omega}}:\mathcal{H}^2\to\mathcal{H}^1$ defined by
$(M_{\sqrt{\omega}} \phi)(p):=\sqrt{\omega}\phi(p)$ (multiplication
operator by $\sqrt{\omega}$) is unitary. The usual position operator
on $\mathcal{H}^2$ is $i\hbar\frac{\partial}{\partial p}$, and is
self-adjoint in some adequate domain. Its counterpart in
$\mathcal{H}^1$ is
$q:=M_{\sqrt{\omega}}\left(i\hbar\frac{\partial}{\partial
    p}\right)M_{\sqrt{\omega}}^{-1}$, the so-called Newton-Wigner
position operator (at time zero) \cite{newton-wigner,lqp}. $q$ is
also self-adjoint, since $M_{\sqrt{\omega}}$ is unitary and $
(q \phi)(p) := i \hbar \left( \frac{\partial}{\partial p}
  - \frac{p}{2 \omega^2} \right) \phi(p) $. The Newton-Wigner
position operator at time $t$, denoted here by $q_t$, is given by
$q_t:=\textrm{e}^{i \omega t /\hbar} q \textrm{e}^{- i \omega t
  /\hbar}$, for $t\in\R$, i.e.,
\begin{equation}
(q_t \phi)(p) 
\; = \; 
 i \hbar \textrm{e}^{ i \omega t /\hbar} \sqrt{\omega}
 \frac{\partial}{\partial p} 
\left( \frac{\textrm{e}^{- i \omega t /\hbar}\phi(p) }{\sqrt{\omega}}\right)
\; = \; 
i \hbar 
\left( \frac{\partial}{\partial p} - \frac{p}{2 \omega^2} - \frac{itp}{\hbar\omega}\right)
 \phi(p) 
\, ,
\label{definicao-qt}
\end{equation}
and is also self-adjoint in $\mathcal{H}^1$. The momentum operator
both in $\mathcal{H}^1$ and $\mathcal{H}^1$ is just multiplication by
$p$.

Recall that the Newton-Wigner states \cite{newton-wigner} localized at
$x$ at time $t$, denoted here by $\psi_{(x, \,t)}$, are given in
momentum representation in $\mathcal{H}^1$ by $\scalar{p}{\psi_{(x,
    \,t)}}\equiv\psi_{(x, \,t)}(p)=\sqrt{\omega/2\pi}e^{i\omega
  t/\hbar-ipx/\hbar}$. They are infinite norm eigenstates of the
Newton-Wigner operator at time $t$ given in \eqref{definicao-qt},
i.e., $q_t\psi_{(x, \,t)}=x\psi_{(x, \,t)}$, what allows to express
$q_t$ in its spectral representation form in $\mathcal{H}^1$ as $ q_t
= \int_{\R} x \ket{\psi_{(x, \,t)}}\bra{\psi_{(x, \,t)}} \;dx
$. Notice that $\int_{\R} \ket{\psi_{(x, \,t)}}\bra{\psi_{(x,
    \,t)}} \;dx$ is formally the identity operator on $\mathcal{H}^1$
and that $\scalar{\psi_{(x, \,t)}}{\psi_{(x', \,t)}}=\delta(x-x')$, as
one easyly checks. The Newton-Wigner wavefunction \cite{lqp}
associated to a state $\phi$ is given by
\begin{equation}
\phi^{NW}(x,\;t) 
\, := \, 
\scalar{\psi_{(x,\,t)}}{\phi}
\, = \,  
\frac{1}{\sqrt{2 \pi}} 
\int_{\R} \frac{\textrm{d}p}{\omega} 
\textrm{e}^{-i\omega t / \hbar + ipx/\hbar } \, \sqrt{\omega} \, \phi(p)  \, .
\label{TheNewtonWignerWavefunction}
\end{equation}
For each $t$, one has $(q_t \phi)^{NW}(x, \,t) = x \phi^{NW}(x,\,
t)$. Therefore, we can interpret the space of the Newton-Wigner
wavefunctions $\phi^{NW}(x,\, t)$ as the spectral representation space
of $q_t$, i.e., the space where it acts as a multiplication operator
\cite{Reed-Simon-1}. In other words, the wavefunction $\phi^{NW}(x,\,
t)$ is the description of the state $\phi$ in the spectral
representation of the Newton-Wigner operator $q_t$. Notice also that $
\scalar{\phi}{\phi'} = \int_{\R} \overline{ \phi^{NW}(x, \, t)
} \, \phi'^{NW}(x, \, t)\;dx $, and we are allowed to regard
$\left|\phi^{NW}(x, \, t)\right|^2$ as the probability density to find
a particle at the position $x$ at the instant $t$.

With these ingredients we can now define localized versions of the
position operator $q_t$ associated to (measurable) regions $O\subset
\R$ by the spectral representation $q_{t,\, O}:=\int_{O} x
\ket{\psi_{(x, \,t)}}\bra{\psi_{(x, \,t)}} \,dx$ or, equivalently, by
\begin{equation}
(q_{t,\,O}\, \phi)^{NW}(x,\;t) \; :=\;  x \, \chi_O(x) \, \phi^{NW}(x,\;t)   \, ,
\end{equation}
where $\chi_O(x)$ is the characteristic function of $O$. The operator
$q_{t,\, O}$ is self-adjoint and is bounded for bounded $O$. 

We now turn to the definition of the coherent states introduced by
Kaiser in \cite{kaiser} and discuss their more relevant properties.
For each $z := (\xi - i \boldpi, \; \tau -
i \boldepsilon) \in \C^2$, with $\xi, \; \boldpi, \, \tau, \, \boldepsilon
\in\R$ and $\boldepsilon > |\boldpi|$, define a coherent state by
\begin{equation}
\phi_z(p) 
\;:=\;
 N^{-1} \sqrt{\omega} 
\exp\left[ 
  i \frac{\omega}{\hbar} (\tau + i \boldepsilon) - i \frac{p}{\hbar} (\xi + i \boldpi)
\right] 
\, .
\label{coherent-def}
\end{equation}
The states originally considered by Kaiser differ from those in
\eqref{coherent-def} in that the factor $\sqrt{\omega}$ is absent in
his formulation. This change corresponds to taking the original states
by Kaiser as elements of $\mathcal{H}^2$, with the states in
\eqref{coherent-def} being the corresponding elements in 
  $\mathcal{H}^1$ obtained by applying the unitary map
  $M_{\sqrt{\omega}}$. More comments about this are found below.  The
normalization constant $N$ is fixed by the condition
\begin{equation}
1 \;=\; \scalar{\phi_z}{\phi_z}
\;=\; 
\int_{\R} \frac{\textrm{d}
  p}{\omega} \left|\phi_z(p)\right|^2
 \;=\; 
\frac{1}{N^2} \int_{\R} \textrm{d} p \, \textrm{e}^{-2(\omega  \boldepsilon - p \boldpi)/ \hbar} 
\;=\;  
\frac{2 m \boldepsilon}{N^2 \lambda} K_1(2m \lambda / \hbar) \, ,
\end{equation}
with $\lambda \equiv \sqrt{\boldepsilon^2 - \boldpi^2}$, where $K_\nu$, here
and below, are the modified Bessel functions of $\nu$-th order
(MacDonald's functions). This leads to $N=\sqrt{2 m \boldepsilon K_1(2m
  \lambda / \hbar) / \lambda}$.  Momentum and position averages can be
calculated in explicit form.  

Their expectation values of $p$ and $q_t$ in the states $\phi_z$ are
given by
\begin{eqnarray}
\average{p}_{\phi_z} & = & m \frac{\boldpi}{\lambda} \frac{K_2(2m\lambda / \hbar)}{K_1(2m\lambda / \hbar)} \, , 
\label{p-rel}   \\
\average{q_t}_{\phi_z} & = & \xi + v (t - \tau) \, , \qquad \mbox{ for } v \; \equiv \frac{\boldpi}{\boldepsilon} \, .
\label{q-rel}
\end{eqnarray}
From the expressions above we see that the average position of the
wavefunction moves with a constant velocity $v$ determined by the
parameters $\boldepsilon,\, \boldpi$, which also determine the average
momentum of the state. The parameter $\xi$ is the initial position of
the coherent state labeled by $z$, and $\tau$ the initial instant of
time. The usual relativistic relation between momentum and velocity
is obtained in the $\hbar \rightarrow 0$ limit, as we will show later.

It is interesting at this point to discuss the localization properties
of the solution of the Klein-Gordon equation associated to the
coherent states $\phi_z(p)$. We will denote these solutions by
$\varphi_z(x,\;t)$. They are given by the Fourier transform
\begin{equation}
\varphi_z(x,\;t) 
 \; = \; 
\frac{2}{\sqrt{2 \pi}} \int_{\R^2} \textrm{d}p_0\textrm{d}p\,  
\textrm{e}^{-i(p_0t - px)/\hbar} \phi_z(p_0,\,p)\theta(p_0) \delta(\mathbf{p}^2-m^2)
\; = \;  
\frac{1}{\sqrt{2 \pi}} \int_{\R} \frac{\textrm{d}p}{\omega} \textrm{e}^{-i\omega t / \hbar + ipx/\hbar } \phi_z(p)  \, .
\end{equation}
This function depends on $x$ and $t$ only through the combinations
$x-\xi$ and $t - \tau$. Hence, for simplicity, we set $\tau = \xi =
0$ and define $z_0\equiv (-i\boldpi, \; -i\boldepsilon)$. 
We are interested in the asymptotic behavior of this wavefunction
when $x \rightarrow \infty$. In order to find it, write
\begin{equation}
\varphi_{z_0}(x,\;t)
\; = \;
\int_{\R} \textrm{d}p\; \textrm{e}^{i p x / \hbar} \left\{ \frac{1}{\sqrt{2 \pi}} \omega^{-1/2} \right\} \left\{ \frac{1}{N} \exp\left[ -\frac{\omega}{\hbar}  (\boldepsilon + i t) + \frac{p \boldpi}{\hbar} \right] \right\}  
\end{equation}
and consider each of the two factors in curly brackets in the integrand
independently, so that the transform can be computed by the convolution
theorem. The Fourier transform of the first factor is just the
Newton-Wigner state localized at $x=0$,
\begin{equation}
\int_{\R} \textrm{d}p\; \textrm{e}^{i p x / \hbar} \frac{1}{\sqrt{2 \pi}} \omega^{-1/2}
\, \propto \,
\left( \frac{2m \hbar}{|x|} \right)^{1/4} K_{1/4}\left( - m|x| / \hbar \right) \, .
\end{equation}
The second factor, after changing to hyperbolic coordinates $ p = m
\sinh s$, $\omega = m \cosh s$, and substituting
\begin{equation}
\boldepsilon + i t \;=\; \rho \cosh y 
\, , \qquad 
\boldpi + ix \;=\; \rho \sinh y \, ,
\label{y-sub}
\end{equation}
with $\Re(\rho) > 0$ and $\Im(y) \in [-\pi/2, \, \pi/2]$, becomes
\begin{equation}
\frac{m}{N} \int_{-\infty}^\infty \textrm{d}s \exp\left[ -\frac{m \rho}{\hbar} \cosh(s-y) \right] \cosh s  \;=\; \frac{2m}{N} \frac{\boldepsilon + i t}{\rho} K_1(- m \rho / \hbar) \, .
\label{position-int}
\end{equation}
Thus,
\begin{equation}
\varphi_{z_0}(x,\;t)
\, \propto \,
\int_{-\infty}^\infty \textrm{d}u \frac{\boldepsilon + i t}{\rho} K_1(-m \rho / \hbar) \, \left( \frac{1}{|x - u|} \right)^{1/4} K_{1/4}\left( - m|x - u| / \hbar \right) \, ,
\label{convolution}
\end{equation}
where $\rho$ is calculated with $u$ replacing $x$ in
\eqref{y-sub}. But $\rho \simeq u + i \boldpi$ for large $u$, and the
MacDonald functions $K_{\nu}$ display an exponential decay for large
arguments. This can be used to prove that the overlap integral in
\eqref{convolution} has an exponential decay for large values of $x$,
i.e., that
\begin{equation}
|\varphi_{z_0}(x,\;t)| \leq \kappa \textrm{e}^{-m |x| / \hbar} \, , \qquad |x| > R \, ,
\label{asymptotic-x}
\end{equation}
for some positive $\kappa,\; R$. This shows that the coherent state
$\phi_z$ is in fact a wavepacket, localized in some finite region,
outside of which it falls to zero exponentially with a mass-dependent
rate. This asymptotic behavior is characteristic of localized
relativistic particles. The Newton-Wigner states satisfy exactly the
same inequality \cite{newton-wigner}. Moreover, the notion of
essentially localized state in Algebraic Quantum Field Theory is also
based on an analog inequality \cite{haag-swieca}.

As we mentioned, the states originally considered by Kaiser differ
from those in \eqref{coherent-def} by the factor $\sqrt{\omega}$. The
basic properties of the coherent states are not affected by the
introduction of this factor, since in both cases the result is a
localized wavepacket moving with constant velocity. The advantage in
introducing this factor is that the velocity $v$ acquires a simple
interpretation in terms of the parameters $\boldpi, \, \boldepsilon$, what not
only will be useful when dealing with the classical limit, but also
gives a more direct physical interpretation of these
coefficients. Moreover, the expression of the associated Newton-Wigner
wavefunction is severely simplified, as will be discussed below. The
payoff is that the spacetime wavefunction $\varphi_z(x,t)$ gets more
complicated. The reason for our choice is that in order to define the
relativistic version of local operators it is natural to use the
Newton-Wigner representation, and in consequence of this we will work
mainly in this representation.

For the Newton-Wigner wavefunction associated to the coherent states
$\phi_{z_0}$ one gets by \eqref{TheNewtonWignerWavefunction} the same
integral solved in \eqref{position-int}, so that
\begin{equation}
\phi^{NW}_{z_0}(x,\;t) 
\;=\; 
\frac{1}{\sqrt{2 \pi}} 
\frac{2m}{N} 
\frac{\boldepsilon + i t}{\rho} K_1(m \rho / \hbar) \, ,
\end{equation}
with $\rho$ given by \eqref{y-sub}. The expressions for general $\xi,
\tau$ are obtained with translations. It follows that the asymptotic
behavior is given in the new representation by the same expression
\eqref{asymptotic-x}.

We have proved that the wavefunction decays exponentially with large
spatial arguments, for a fixed time. On the other hand, from
\eqref{q-rel} it is seen that the average position moves with constant
velocity. Now we want to discuss how the wavepacket spreads about this
average motion, so consider the variance of the position
distribution. It turns out that
\begin{equation}
\sigma_{q_t}^2 
\;=\; 
\average{q_t^2}_{\phi_{z_0}} -  \average{q_t}_{\phi_{z_0}}^2 
\;=\; 
- v^2 t^2 - \boldpi^2 
+ (t^2 + \boldepsilon^2) \frac{1}{N} \int_{\R}\textrm{d}p \,
  \frac{p^2}{\omega^2} \textrm{e}^{- 2(\omega \boldepsilon - p \boldpi)/2\hbar} 
\;=:\; 
D(\hbar)^2  \, ,
\end{equation}
and that $\lim_{\hbar \rightarrow 0 } \sigma_{q_t}^2 = 0$ (uniformly
for $t$ in compacts). Therefore, the wavepacket is well concentrated
about the average motion for small $\hbar$. The momentum is also
well-determined in this limit,
\begin{equation}
\sigma_{p}^2 
\;=\; 
\average{p^2}_{\phi_{z_0}} - \average{p}_{\phi_{z_0}}^2 
\;=\; 
\frac{m^4}{4} \frac{K_3(2m\lambda / \hbar) - K_1(2m\lambda /
  \hbar)}{K_1(2m\lambda / \hbar)} 
+ \frac{m^2 \boldpi^2}{\lambda^2} 
 \left[\frac{K_3(2m\lambda / \hbar)}{K_1(2m\lambda / \hbar)} - \left(
    \frac{K_2(2m\lambda / \hbar)}{K_1(2m\lambda / \hbar)} \right)^2  
\right]\, ,
\end{equation}
what leads to $\lim_{\hbar \rightarrow 0} \sigma_{p}^2 = 0$. In this
limit, $p = m \boldpi/ \lambda = m \gamma(v) v$, for
$\gamma(v):=(1-v^2)^{-1/2}$, and the usual relativistic relation
between momentum and velocity is obtained.

The fact that $\lim_{\hbar \rightarrow 0 } \sigma_{q_t}^2 = 0$ can be
used to determine a nice property of the Newton-Wigner wavefunctions
of the coherent states. Let us now take $\xi\neq 0$ but $\tau=0$.  In
the Newton-Wigner representation one has $\sigma_{q_t}^2 =
\int_{\R}\textrm{d}x \, \left| \phi_z^{NW}(x,t) \right|^2
[x-(\xi + vt)]^2 $. For $R>0$, define the time-dependent region $
O_{t,\,\xi}:=\{x;\; |x-(\xi + vt)| < R\}$ with $O_{t,\,\xi}^c$ being
its complementary set. Hence,
$$
R^2 \int_{O_{t,\,\xi}^c}\textrm{d}x \, \left| \phi_z^{NW}(x,t)\right|^2
\; \leq \; 
\int_{O_{t,\,\xi}^c}\textrm{d}x \, \left| \phi_z^{NW}(x,t) \right|^2 [x-(\xi + vt)]^2
\; \leq \;
\sigma_{q_t}^2
\;,
$$
implying that $ \int_{O_{t,\,\xi}^c}\textrm{d}x \, \left|
  \phi_z^{NW}(x,t)\right|^2\leq D(\hbar)^2 / R^2$. Since
$\lim_{\hbar\to 0}D(\hbar)=0$, we may claim that, for each fixed $R$,
the fraction of the $L^2$ norm of $\phi_{z_0}^{NW}(x,\;t)$ outside the
time-dependent region $ O_{t,\,\xi}$ is smaller than any desired bound
$D_0 > 0$ as $\hbar \rightarrow 0$.
This is a relativistic analog of
\eqref{eq:localized-1state-bound}. Besides that, it follows by
analogous computations that the average of the local operator is a
good approximation for the average of $q_t$:
\begin{equation}
\left| \average{q_t}_{\phi_z} - \average{q_{t,\, O_{t,\,\xi}}}_{\phi_z} \right|
\, \leq \,
\frac{D(\hbar)}{R} \left[ |\xi + vt| + D(\hbar) \right] \, .
\end{equation}

Summing up, the situation is very similar to that found in the
nonrelativistic case: coherent states are states essentially localized
in some region $O$, outside of which the wavefunction is as small as
desired when $\hbar \rightarrow 0$, and the average of the position
operator can be approximated by a local version. We will now show that
the same steps followed there can be repeated here, and an
approximation scheme can be devised for the free scalar field which
leads to a classical limit describing a system of relativistic
particles.

First, define $N$-body local operators $q_{t,\,O}^{(N)}$ acting on the
$N$-particle sector of the Fock space generated by $\mathcal{H}^1$ by
the recipe given in \eqref{eq:nbody-local-op-def}. At each time
$t$, essentially localized $N$-particle states $\Phi^{N}$ are defined
as symmetrized products of one-particle coherent states
$\phi_i\equiv\phi_{z_i}$ situated at disjoint regions $O_i = [a_i - R,
\; a_i + R]$, where $a_i = \xi_i + v_i t$. Let $d$ be the smallest
distance between these regions, and put $A := \max\{|a_i|\}$. Then,
\begin{equation}
\left| \scalar{\phi_i}{\phi_j} \right| \; \leq \; \frac{2 D(\hbar)}{R} \, ,
\end{equation}
and for the operator $q_{t,\,O_k}$,
\begin{align}
| \bra{\phi_i} q_{t,\,O_k} \ket{\phi_j} |  & \;\leq\;  (A + R) \, \frac{D(\hbar)}{R+d} \, , \qquad \textrm{if } i\neq j \, ,   	\\
| \langle q_{t,\,O_k} \rangle_{\phi_i} | & \;\leq\;  \frac{A D(\hbar)}{R + d} \, , \qquad \textrm{if } i\neq k \, ,		 \\
|\langle q_t \rangle_{\phi_k} - \langle q_{t,\,O_k} \rangle_{\phi_k} | & \;\leq\;  (A+D(\hbar)) \frac{D(\hbar)}{R}  \, .
\end{align}
These are the analogs of
Eqs.~\eqref{scalar-product-bound}--\eqref{local-op-bounds}. Then, it
turns out that
\[
\left\langle q^{(N)}_{t,\,O_k} \right\rangle _\Phi \;\simeq\; \langle q_t \rangle_{\phi_k} \, ,
\]
up to an error of order $D(\hbar)$. But the average at the right can
be evaluated in the classical limit $\hbar \rightarrow 0$, leading to
\[
\lim_{\hbar \rightarrow 0} \left\langle q^{(N)}_{t,\,O_k} \right\rangle _\Phi 
\;=\; 
\xi_k + v_k t \, .
\]
This completes our argument. The complete result is the following:
there are localized position operators defined in each $N$-particle
sector for the free scalar field whose mean values, when evaluated at
essentially localized states constructed as the symmetrized products
of relativistic coherent states, follow the expected classical
trajectories in the limit $\hbar \rightarrow 0$ .

We have restricted the definition of the local operators to some
$N$-particle sector, but this is not necessary. One can define local
operators $ \bigoplus_{N=1}^\infty q_{t,\,O}^{(N)} $ in Fock space
which measure positions inside a specified region $O$ of space. The
average of this operator for an essentially localized state
$\ket{\psi}$ with any number of localization centers $O_i$ is
approximately zero if $O_i \cap O = \emptyset$ for all $i$, and is
approximated by $\bra{\phi_k} q_t \ket{\phi_k}$ if $O_k = O$ for some
$k$. The number of particles of the state $\ket{\psi}$ is irrelevant;
the local observer at $O$ cannot determine how many particles are
there in regions of space which are not accessible to it.

The above results can be viewed in connection with the question of
characterizing which quantum field theories can be described in terms
of particles. The relativistic coherent states we have discussed
provide a quantum representation for the classical concept of a
relativistic particle in free motion. Whenever one uses a quantum
field theory to describe scattering processes between interacting
particles, it is somehow assumed that states which correspond to
particles do exist. This requirement is usually formulated as the
condition of existence of essentially localized states whose
wavefunction decays exponentially with a mass-dependent coefficient at
any fixed instant of time. This is the basic idea underlying the
approach of \cite{haag-swieca}. If one is willing to be more restrictive,
there is the possibility of imposing the additional condition that
localized states remain localized at all times, and this line of
argument was pursued in \cite{enss}. The discussion above suggests an
alternative approach. One may try to associate coherent states to
particle tracks observed in experiments. These states must be
essentially localized, but besides that, one should impose the
condition that the average of suitable local position operators must
follow classical trajectories, with a negligible dispersion about this
average, at least for a time-interval compatible with the experiment
in question. In this case the question of the existence of a particle
interpretation could be recast as the question of existence of a
classical limit describing particles, and Hepp's analysis would be a
natural tool to deal with this problem.

\section{Conclusion}

\label{sec:conclusion}

We have defined essentially localized states and local versions of
Weyl operators for nonrelativistic bosonic particles, and used these
concepts to prove the existence of a particle classical limit for a
system of $N$ interacting bosonic particles in an external
potential. The quantum theory of a massive scalar field was proved to
have two distinct classical limits, describing classical particles or
a classical field. Local versions of the Newton-Wigner position
operator were defined for this theory, and the particle classical
limit was obtained as the $\hbar \rightarrow 0$ limit of the average
of these local operators evaluated on a class of essentially localized
states constructed as symmetrized products of relativistic coherent
states introduced herein.

N.Y.\ thanks FAPESP for financial support.

\appendix

\section{Time-evolution of coherent states for linearized equations of motion}

\label{time-evolution}

A system of $N$ distinguishable particles of the same mass ($m=1$, for
convenience) is described by the (unsymmetrized) Hilbert space
$\mathcal{H}^{\otimes N}$ of square-integrable functions on
$\R^N$. Consider the time-evolution generated by the
time-dependent quadratic Hamilton operator
\begin{equation}
H(t) 
= 
\sum_{i=1}^N  \frac{p_i^2}{2} 
+ \frac{1}{2} \sum_{i \neq j} V^{\prime \prime}(\xi_i(t) - \xi_j(t)) (q_i - q_j)^2 \, ,
\end{equation}
where the $\xi_i(t)$, $j=1, \, \ldots , \, N$, describe the classical
trajectories of $N$ interacting particles for a solution of the
corresponding classical system with initial conditions $\alpha = [\xi
+ i\pi] / \sqrt{2 \hbar}$. In the Heisenberg picture, the
time-dependent position and momentum operator satisfy linear equations
\[
i \hbar \dot{q}_i(t) 
\, = \,  
p_i(t) 
\, , \qquad 
i \hbar \dot{p}_i(t) 
\, = \, 
- \sum_{i \neq j} V^{\prime \prime}(\xi_i(t) - \xi_j(t)) q_j(t) 
\, ,
\]
identical to the classical equations of motion. These are solved by
\[
\begin{bmatrix} q(t) \\ p(t)
\end{bmatrix} = S(t)
\begin{bmatrix} q \\ p
\end{bmatrix} \, ,
\]
where $S(t)$ is a $(2N)\times(2N)$ symplectic matrix (in order that
the canonical commutation relations be preserved) whose entries
depend continuously on $t$. Writing $S(t)$ in block form as
\[
S(t) = \begin{bmatrix} A(t) & B(t) \\ C(t) & D(t)
\end{bmatrix} \, ,
\]
the symplectic condition is equivalent to $A D^t - B C^t = \mathbb{1},
AB^t = BA^t, CD^t = DC^t$ \cite{arvind}. Here, $A$, $B$, $C$, $D$ and
$\mathbb{1}$ are $N\times N$ matrices, $\mathbb{1}$ being the identity
matrix.

Now, let the initial state of the system be the coherent state
$\ket{0}$ of zero position and momentum in $\mathcal{H}^{\otimes N}$,
and $\psi_0(x)=\scalar{x}{0}$ be the corresponding wavefunction. The time-evolved
state at instant $t$ in the Schr\"odinger representation is 
$\psi(x,\, t) = \bra{x} W(t) \ket{0}$ as usual, where $W(t)= \text{T}
\exp \left[ - i / \hbar \int_0^t \text{d}t^\prime H^\prime(t^\prime)
\right]$. The initial state satisfies the differential equations
\begin{equation}
[q_j + i p_j] \psi_0(x) = 0  \, , \qquad j = 1,\, \dotsc,\, N \,,
\label{eq:psi-0}
\end{equation}
what in turn implies that $[q_j(t) + i p_j(t)] \psi(x,t) = 0$,
$\forall j$. In terms of the block-components of $S(t)$, one has
\[
\left[ 
 \bigl( A^r_s(t) + i C^r_s(t) \bigr) x_s - \hbar \bigl( D^r_s(t) - i B^r_s(t) \bigr) \nabla_s 
\right] \psi(x,\, t) 
\; = \; 0 
\, .
\]
The indices $r,\, s=1,\, \ldots ,\, N$ label matrix coefficients and
the Einstein sum convention is used. Some results about symplectic
matrices allow for an exact solution of this linear system of
differential equations. First, it is advantageous to break $S$ into
simpler factors, and solve successively for the action of each factor.
It was proved in \cite{xu} that a so-called structured singular value
decomposition exists for any symplectic matrix $S$: one can write
\begin{equation}
S = O^\prime D O \, ,
\label{eq:svd}
\end{equation}
where $O,\, O^\prime$ are orthogonal, $D = \text{diag}\big(\omega_1,
\, \dotsc,\, \omega_N,\, \omega_1^{-1},\, \dotsc,\,
\omega_N^{-1}\big)$, with $\omega_j > 0$ for all $j$, and the three
factors in the decomposition are symplectic matrices. That any square
matrix $M$ admits a singular value decomposition (i.e., $M$ can be
written as $M=O_1D_1O_2$, with $O_1$ and $O_2$ orthogonal and $D_1$
diagonal with non-negative entries) is a well-known result; a
structured decomposition is one in each all three factors are symplectic.
Second, an orthogonal symplectic matrix $O$ has a very simple form,
\begin{equation}
O = 	\begin{bmatrix} U & V \\ - V & U
	\end{bmatrix} \, ,
\label{eq:orthogonal}
\end{equation}
where $U,\, V$ are real matrices such that $U-iV$ is unitary.

Hence, for a given time $t$ with $|t| < T$ write $S(t)$ in the form of
a singular value decomposition \eqref{eq:svd}, with $O$ as in
\eqref{eq:orthogonal}. Consider first the action of the factor $O$.
One can define new operators
\[
\begin{bmatrix} q^\prime \\ p^\prime
\end{bmatrix} = O 
\begin{bmatrix} q \\ p
\end{bmatrix}
\]
which satisfy the canonical commutation relations $[q_j^\prime,\,
p_k^\prime] = i \hbar \delta_{jk}$. But, by the Stone-von Neumann
Theorem, there is, up to unitary equivalence, only one representation
of the canonical commutation relations\footnote{Strictly speaking, the
  uniqueness statement of the S.-von N.\ Theorem refers to the
  representations of the Weyl form of the canonical commutation
  relations.}, hence the pair $q^\prime,\, p^\prime$ must be unitarily
equivalent to the original pair $q,\, p$.  Let $U_O$ be a unitary
operator such that $q^\prime_j = U_O^* q_j U_O$, and $p^\prime_j =
U_O^* p_j U_O$. Use this operator to define the state $\psi^\prime(x)
= U_O \psi(x)$ which, from \eqref{eq:psi-0}, must satisfy
\[
(U^r_s - i V^r_s) (x_s + \hbar \nabla_s) \psi^\prime(x) 
\; = \;  
0 
\; .
\]
Multiplying by $U^t + i V^t$ on the left, one gets $(x_s + \hbar
\nabla_s) \psi^O(x) = 0$, whose solution is $\psi^\prime(x) =
\psi(x)$. The initial state is not changed by the action of $O$. The
action of the factors $D$ and $O^\prime$ can be studied in the same manner.
Define 
\[
\begin{bmatrix} q^{\prime \prime} \\ p^{\prime \prime}
\end{bmatrix} 
\; := \; 
D \begin{bmatrix} q^\prime \\ p^\prime
  \end{bmatrix} 
\, , \qquad 
\begin{bmatrix} q^{\prime \prime \prime} \\ p^{\prime \prime \prime}
\end{bmatrix} 
\; := \; 
O^\prime \begin{bmatrix} q^{\prime \prime} \\ p^{\prime \prime}
        \end{bmatrix} 
\, ,
\]
and let $U_D,\, U_{O^\prime}$ be unitary operators which accomplish the
corresponding unitary transformations. The state $\psi^{\prime \prime}
\equiv U_D \psi$ satisfies the system of differential equations
\[
(\omega_r x_r + \hbar \omega_r^{-1} \nabla_r) \psi^{\prime \prime}(x) 
= 
0 
\, ,
\]
which is solved by the product of Gaussian functions $\psi^{\prime
  \prime}(x) = \prod_j (\pi \hbar)^{-1/4} \omega_j^{1/2} \exp[-
(\omega_j x_j)^2 / 2 \hbar]$. This state is not changed by the action
of $O^\prime$, so that $U_{O^\prime} U_D U_O \psi(x) = \psi^{\prime
  \prime}(x)$. But $U_{O^\prime} U_D U_O = U_S$, and the unitary
operator which correspond to the symplectic matrix $S$ is just the
propagator $W(t)$. Therefore,
\[
\psi(x,\, t) 
\; = \;  
\prod_j (\pi \hbar)^{-1/4} \omega_j^{1/2} \exp[- (\omega_j x_j)^2 / 2 \hbar] 
\, .
\]
The time-evolution of a coherent state under a quadratic Hamiltonian
can always be solved via a singular value decomposition. An algorithm
giving the decomposition of an arbitrary symplectic matrix was
developed in \cite{xu}.

\end{document}